\documentclass[doublecol]{epl2}

\title{Bipolaron formation in organic semiconductors at the interface with dielectric gates}

\author{C. A. Perroni\inst{1} and V. Cataudella\inst{1}}

\institute{
  \inst{1} CNR-SPIN and Dipartimento di Scienze Fisiche, Univ. di Napoli ``Federico II'', I-80126 Italy \\
}

\abstract{The formation of the electron-phonon induced bipolaron is shown to be feasible in organic semiconductors at the interface with dielectric gates due to the coupling of the carriers with interface vibrational modes and to the weak to intermediate strength of bulk electron-electron interaction. The polaronic bound states are found to be quite robust in a model with realistic strengths of electron coupling to both bulk and interface phonons. The crossover to nearly on-site bipolarons occurs for coupling values much smaller than those for nearly on-site polarons, but, on the other hand, it gives rise to an activated behavior of mobility with much larger activation energies. The results are discussed in connection with rubrene field-effect transistors.
}

\begin{document}

\maketitle

{\it Introduction}
In the last years, field-effect devices have been realized with correlated oxides and organic materials. ~\cite{RMP}
This has allowed to inject charge carriers into complex materials inducing new phases and properties. On the other hand, the devices work with solid or liquid dielectric gates and new effects has to be taken into account: for example, the lattice mismatch between bulk and gate, the dielectric breakdown for strong electric fields, and the interaction of bulk electrons with gate degrees of freedom.

Single crystal organic field-effect transistors (OFET) are important not only for applications, but also for the understanding of the intrinsic properties of organic semiconductors (OS). \cite{hasegawa} The most promising and studied systems are those realized with oligoacenes, such as rubrene. \cite{morpurgo} In the OFET geometry, the nearby dielectric can have a strong influence on the transport properties. Indeed, some experiments have pointed out the existence of scaling laws of the mobility as a function of the dielectric constant of solid \cite{stassen,nature} and liquid \cite{ono} gates. If the difference between the static and high frequency dielectric constant of the gate is small (for example for parylene), the mobility $\mu$ of these systems exhibits a power-law behavior ($\mu \sim T^{-\gamma}$, with $\gamma \simeq 2$)  at temperatures close or higher than $100 K$. \cite{morpurgo} This conduction process cannot be simply ascribed to band transport,\cite{cheng,corop,hannewald,troisi_prl} and it is believed to arise from the interaction of charge carriers with bulk modes. \cite{troisi_prl,vittoriocheck}  On the other hand, if the dielectric constant mismatch is high (for example for $Ta_2 O_5$), an insulating behavior is found much below room temperature with a strong decrease of the mobility.

The explanation of these behaviors has been based on the possibility that the injected charge carriers undergo a polaronic localization due to the interaction with the polarizable gate dielectrics, leading to the formation of Frohlich polarons. \cite{nature,fratini,bussac,bussac2} However, the predicted values of the activation energy are smaller than those estimated by accurate calculations. \cite{bussac1} For example, the barrier of rubrene transistors with a $Ta_2 O_5$ oxide gate is calculated to be $12.4$ meV, a value much lower than the energy of $55$ meV found in previous estimates. \cite{nature} For this reason, recently, the interplay between charge carrier coupling with bulk vibrations and the long-range interaction induced at the interface with the dielectric gate has been investigated. \cite{giulio} These interactions stabilize a polaronic state with specific properties, that, however, requires values of the bulk electron-phonon (el-ph) coupling still larger than those typically estimated. One way to solve this inconsistency is to include the effect of disorder.\cite{morpurgo,single}

Recently, growth and purification techniques of OS are ever and ever improving tending to strongly reduce the amount of traps in the bulk and at the interfaces. \cite{liu} Therefore, in this work, we focus on the clean limit proposing that the bipolaron formation induced by the long-range interaction of the carriers with interface vibrational modes can affect the transport properties. The bipolaron mechanism has been invoked in organic single molecules \cite{geskin} and disordered systems. \cite{bobbert,yaping} However, bound polaronic states have not carefully investigated in clean single-crystal OS. The polaronic bound states are studied in a model, which, in addition to bulk and interface el-ph couplings, includes local Hubbard and non local long-range Coulomb interactions treated through exact diagonalizations. The bipolaron formation is possible due to the weak to intermediate strength of the bulk electron-electron interaction in OS.

In spite of the effects of bulk el-ph interactions, the formation of bound states takes place for realistic strengths of the interface el-ph coupling. With increasing the interface interaction, we point out that the bound states form with different features: large bipolarons with extended lattice distortions for weak to intermediate coupling, small bipolarons with essentially on-site distortions for intermediate to strong coupling. In particular, the crossover to small bipolarons occurs for coupling values much smaller than those for the small polaron formation. On the other hand, we will show that the conduction process of small bipolarons is characterized by a hopping behavior with activation energies much higher than those of the single small polaron, in qualitative agreement with data in rubrene OFET.

{\it Model}
We study the following Hamiltonian model  with coupling to bulk and interface vibrational modes \cite{giulio}, local Hubbard and non-local long-range Coulomb interactions:
\begin{equation}
H= H_{el}+H_{B}^{(0)}+H_{el-B}+H_{I}^{(0)}+H_{el-I}.
\label{h}
\end{equation}

In Eq. (\ref{h}), the electronic part $H_{el}$ is
\begin{eqnarray}
H_{el}&=& -t \sum_{i,\sigma}  \left( c_{i,\sigma}^{\dagger} c_{i+1,\sigma}+ h.c. \right) +U \sum_{i}
n_{i,\uparrow} n_{i,\downarrow} + \nonumber\\
&&  \frac{e^2}{\epsilon} \sum_{i \ne j } \frac{e^{-\eta |R_i-R_j|}}{|R_i-R_j|} n_{i} n_{j}.
\label{hel}
\end{eqnarray}
The kinetic energy operator is in the first term of Eq. (\ref{hel}), in which  $t$ is the bare electron hopping between the nearest neighbors on a chain with lattice parameter $a$, $c_{i,\sigma}^{\dagger}$ and $c_{i,\sigma}$ are the charge carrier creation and annihilation operators, respectively, relative to the site $R_i$ and spin $\sigma$. The {\it ab-initio} estimate for $t$ is around $50-100 meV$. \cite{corop} Within the model, one considers a one-dimensional stack of planar conjugated molecules (axis with the highest hopping integral). \cite{troisi_prl}
The local Hubbard interaction with strength U is in the second term of Eq. (\ref{hel}), in which the density operator of spin $\sigma$ electrons is $n_{i,\sigma}=  c_{i,\sigma}^{\dagger} c_{i,\sigma} $. OS are not strongly correlated, therefore the quantity U is estimated to be less than or of the order of half-bandwidth. \cite{bobbert}
The non local Coulomb interaction is in the third term of Eq. (\ref{hel}), in which $e$ is the electron charge, $\epsilon$ is the bulk relative dielectric constant, $\eta$ is a cut-off distance for long-range interactions on finite lattices, and $n_{i} =  \sum_{\sigma} n_{i,\sigma}$. The reference energy for the Coulomb interaction is $V=e^2 / \epsilon a$, with $\epsilon$ between $2$ and $3$ for OS.

In Eq. (\ref{h}), $H_{B}^{(0)}$ is Hamiltonian of free bulk modes. {\it For rubrene, the mixing of low frequency intramolecular modes with intermolecular slightly dispersive lattice phonons takes place. \cite{girlando} Therefore, we consider a single average dispersion-less mode, \cite{troisi_adv,newus} implying}
\begin{equation}
H_{B}^{(0)}= \sum_{i} \frac{{p}^2_{i}}{2 m}+ \sum_{i} \frac{ k x_{i}^{2}}{2},
\label{hintra}
\end{equation}
where $x_{i}$ and ${p}_{i}$ are the $i-th$ molecule displacement transverse to the chain and the momentum, respectively, $m$ the oscillator mass, $k$ the elastic constant and $\omega_{B}=\sqrt{k/m}$ the mode frequency. These bulk modes are characterized by small energies ($\hbar \omega_{B} \simeq 5-10$ meV, with $\hbar$ Planck constant). \cite{corop,troisi_prl}

In Eq. (\ref{h}),  $H_{el-B}$ represents the term similar to the Su-Schrieffer-Heeger \cite{SSH} (SSH) interaction for the coupling to bulk modes
\begin{equation}
H_{el-B}= \alpha \sum_{i,\sigma} (x_{i+1}-x_i) \left( c_{i,\sigma}^{\dagger}c_{i+1,\sigma}+ h.c.  \right),
\label{hcoupling1}
\end{equation}
with $\alpha$ coupling constant. \cite{troisi_prl,vittoriocheck} This coupling induces a modulation of the transfer integral between two neighbor molecules. In the adiabatic regime for bulk modes ($\hbar \omega_{B} \ll t$), the dimensionless quantity $\lambda_{B}=\alpha^2/4 k t $ fully provides the strength of the bulk el-ph coupling. The typical values of $\lambda_{B}$ are in the intermediate regime (in this work, we take $\lambda_{B}=0.1$ appropriate for rubrene \cite{vittoriocheck}).

In Eq. (\ref{h}), $H_{I}^{(0)}$ is the Hamiltonian of free interface phonons
\begin{equation}
H_{I}^{(0)}= \hbar \omega_{I} \sum_{q} a_q^{\dagger} a_q,
\label{hintra}
\end{equation}
where $\omega_{I}$ is the frequency of optical modes, $a_{q}^{\dagger}$ and $a_{q}$ are creation and annihilation operators, respectively, relative to phonons with momentum $q$. In comparison with models in high dimensions, \cite{bussac} interface modes are assumed purely one-dimensional propagating along the electron chain lying close to the gate. \cite{giulio}

In Eq. (\ref{h}), $H_{el-I}$ is the Hamiltonian describing the electron long-range coupling to interface vibrational modes
\begin{equation}
H_{el-I}= \sum_{j,q} M_q n_j e^{i q R_j} \left( a_q + a_{-q}^{\dagger}   \right),
\label{hcoupling}
\end{equation}
where $M_q$ is interaction el-ph term
\begin{equation}
M_q= \frac{g \hbar \omega_{Inter} }{\sqrt{L}} \sum_{j} e^{i q R_j} \frac{R_0^2}{R_0^2+R_j^2},
\label{hcoupling}
\end{equation}
with $g$ dimensionless coupling constant, $L$ number of lattice sites, and $R_0$ cut-off length of the order of the lattice spacing $a$. The length $R_0$ indicates the distance of the chain from the polar gate. In this work, we take $R_0=0.5 a$ and $\hbar \omega_{Inter}=0.5 t$. \cite{fratini,bussac}
In order to quantify this coupling, we use the dimensionless quantity $\lambda_I=\sum_q M_q^2/2 \omega_{I} t$. We stress that, when one compares this coupling with the more realistic two-dimensonal case, one finds a similar long-range behavior. \cite{giulio}

In the following, we will use units such that $a=1$, $\hbar=1$, $e=1$, and Boltzmann constant $k_B=1$.
We will analyze bipolaron systems in thermodynamic limit (up to $L=32$ for $\eta=0.1$) measuring energies in units of
$t \simeq 80 meV$. \cite{nota} We fix $U=2.0 t$. \cite{bobbert}

{\it Method}
The temperature range relevant for transport properties is such that $\omega_{Bulk} < T$, where the adiabatic approximation for bulk lattice degrees can be considered valid. \cite{vittoriocheck} Therefore, the dynamics of intermolecular bulk modes can be treated as classical. The electron dynamics is strongly influenced by the lattice oscillations which are described by the distribution function $P \left( \{ x_i \}  \right) $, which is self-consistently calculated through a Monte-Carlo approach. \cite{vittoriocheck}
At a fixed configuration of bulk displacements $\{ x_i \}$, Eq.(\ref{hcoupling1}) shows that the electron hopping between nearest neighbor sites is not homogeneous and depend on the specific pair.

At a fixed configuration $\{ x_l \}$, the resulting inhomogeneous model can be accurately diagonalized using the following basis states
\begin{equation}
U_I \left( \{ x_l \}  \right) |n_{q_1},.,n_{q_L} > |i,j>_{S_z=0},
\label{basis}
\end{equation}
where the unitary transformation $U \left( \{ x_l \} \right)$
\begin{equation}
U \left( \{ x_l \}  \right)= \exp{ \left[ \frac{1}{\sqrt{L}} \sum_{j,q} f_q\left( \{ x_l \} \right) n_j e^{i q R_j} \left( a_q - a_{-q}^{\dagger}   \right) \right] } ,
\end{equation}
provides an el-ph treatment equivalent to a generalized Lang-Firsov approach. \cite{lang} For long-range el-ph interactions, this approach is very accurate in the intermediate to strong coupling regime. \cite{alexandrov} The parameters $f_q\left( \{ x_l \} \right)$ can be self-consistently calculated or, without losing accuracy, parametrized as $f_q \left( \{ x_l \} \right)=a\left( \{ x_l \} \right) (M_q/ \omega_{Inter})/(b\left( \{ x_l \} \right) \cos (q)+1)$, with $a \left( \{ x_l \} \right)$ and $b \left( \{ x_l \} \right)$ configuration dependent variational quantities. \cite{vitmanga} We point out that, after the diagonalization at fixed displacements $\{ x_l \}$, the search for the best variational parameters $f_q\left( \{ x_l \} \right)$ is made through the minimization of the free energy F. This procedure is based on intensive exact diagonalizations. Through the variational parameters $f_q\left( \{ x_l \} \right)$, interface phonons are displaced from their equilibrium position to a distance proportional to the el-ph interaction.
In Eq. (\ref{basis}), $|n_{q_1},.,n_{q_L} >$ denotes the phononic basis in the occupation number representation, and $|i,j>_{S_z=0}=c_{i,\uparrow}^{\dagger}c_{j,\downarrow}^{\dagger}|0>$, with $|0>$ vacuum state, is the basis state for the electron pair with total spin component $S_z=0$. We consider this Hilbert subspace since, for large el-ph coupling, the bipolaron is in a singlet state ($S=S_z=0$).

In order to clarify the nature of the obtained solutions, some correlation functions such as the double occupation $d=< \sum_i n_{i,\uparrow} n_{i,\downarrow} >$ and the nearest neighbor quantity
$p=< \sum_i n_{i,\uparrow} n_{i+1,\downarrow} >$ are calculated. The main aim of this work is the mobility as a function of the temperature.
The mobility $\mu \left( \{ x_l \} \right)$ is determined starting from the real part of the conductivity
$Re[\sigma(\omega)] \left( \{ x_l \} \right)$, taking the limit of small frequency and dividing for the particle density $2/L$ of the system.
In the linear regime, the real part of the conductivity is derived from the Kubo formula \cite{mahan}
\begin{equation}
Re[\sigma(\omega)] \left( \{ x_l \} \right)=\frac{ \left( 1-e^{-\beta \omega} \right) }{2 \omega} \frac{1}{L} \int_{-\infty}^{\infty} d t e^{i \omega t}
\langle j^{\dagger}(t) j(0) \rangle,
\label{kubo}
\end{equation}
where
\begin{equation}
j=-i \sum_{j,\delta,\sigma} \delta c^{\dagger}_{j+\delta,\sigma} c_{j,\sigma} h_{j,\delta} \left( \{ x_l \} \right)
\end{equation}
is the current operator, $ h_{j,\delta} \left( \{ x_l \} \right)= t-\alpha \delta (x_{j+\delta}-x_j) $ denotes the generalized hopping and $\delta=-1,1$ indicates the nearest neighbors.

If an observable $O$ depends on the displacements $\{ x_i \}$, first, one makes the average $ \left\langle O \right\rangle \left( \{ x_i \} \right) $ over the eigenstates and eigenvectors of the corresponding inhomogeneous model, then over the distribution $P \left( \{ x_i \} \right)$ making the integral
\begin{equation}
\left\langle \left\langle O \right\rangle \right\rangle =\int  \left( \prod_{i} d x_{i} \right)    P \left( \{ x_i \} \right)
\left\langle O \right\rangle \left( \{ x_i \} \right)
\label{distri}
\end{equation}
by means of the Monte-Carlo procedure. \cite{vittoriocheck}

\begin{figure}[htb]
\flushleft
\onefigure[width=0.42\textwidth,height=0.52\textwidth,angle=0]{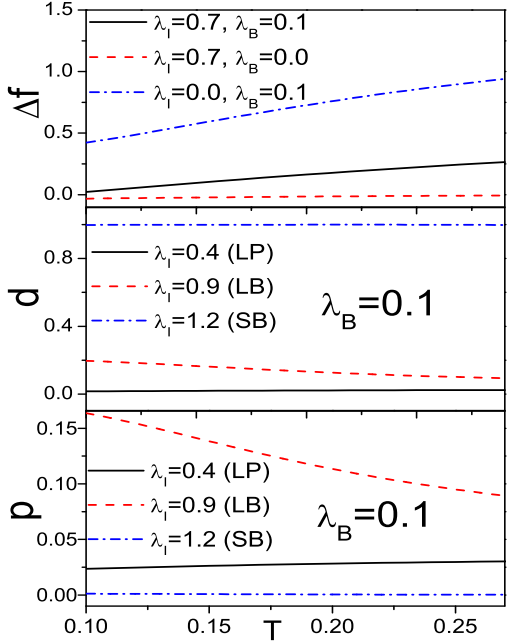}
\caption{The free energy difference $\Delta f$ (in units of $t$), the double occupation $d$ and the nearest neighbor correlation function $p$ as a function of the temperature (in units of $t$) for different values of interface ($\lambda_{I}$) and bulk ($\lambda_{B}$) el-ph couplings. (LP) stands for large polaron, (LB) large bipolaron, and (SB) small bipolaron.}
\label{spectral1}
\end{figure}

{\it Results}
First, we analyze the stability of bipolarons comparing the free energy of the two-particle problem $F_{2P}$ with that of two single polarons $F_P$. The bipolaron is stable if the difference $\Delta f=F_{2P}-2 F_P <0$. In the upper panel of Fig. 1, we report the free energy difference $\Delta f$ for different values of interface and bulk el-ph couplings as a function of the temperature. By means of the correlation functions $d$ and $p$, we can follow the crossover between large polaron (LP), large bipolaron (LB), and small bipolaron (SB) states. The adjective large means extended range of lattice distortions, which implies that, in $f_q$, $a$ is smaller than unity and  $b$ is different from zero. The adjective small refers to local lattice distortions, and corresponds to $a=1$ and $b=0$. In Fig. \ref{spectral1}, we show that the SB is characterized by the double occupation $d=1$ and the nearest neighbor correlation $p=0$ pointing out that the SB is made of two particles with spin up and spin down on the same site. On the other hand, the LB solution exhibits a moderate value of $d$ and $p$, both decreasing with temperature. As expected, the wave-function of LB spreads out over many sites, and the spreading gets larger with increasing temperature. \cite{alexandrov}

Surprisingly, bulk el-ph coupling tries to change the structure of the LB and SB solutions reducing their stability. As shown in the upper panel of Fig. \ref{spectral1}, the effect of the bulk coupling ($\lambda_B=0.1$) forbids the formation of bound states at  $\lambda_I=0.7$. { Apparently, the bulk coupling alone is not able to bind the two particles}. The bulk SSH-like interaction reduces the ratio between the particle average kinetic energies with and without e-ph coupling, tending to enhance the localization of the particle and decreasing the free energy $F_P$. With increasing the bulk el-ph interaction, similarly, the electron pair reduces the free energy $F_B$. However, the bulk el-ph coupling tends to extend the spatial correlation of the local SB solution decreasing $d$ and increasing $p$. On the other hand, in the LB regime, the bulk el-ph interaction tends to enhance both $d$ and $p$. Therefore, the bulk interaction tends to decrease the pairing range for the LB going towards bond correlations. \cite{meSSH}

\begin{figure}[htb]
\centering
\onefigure[height=0.37\textwidth,angle=0]{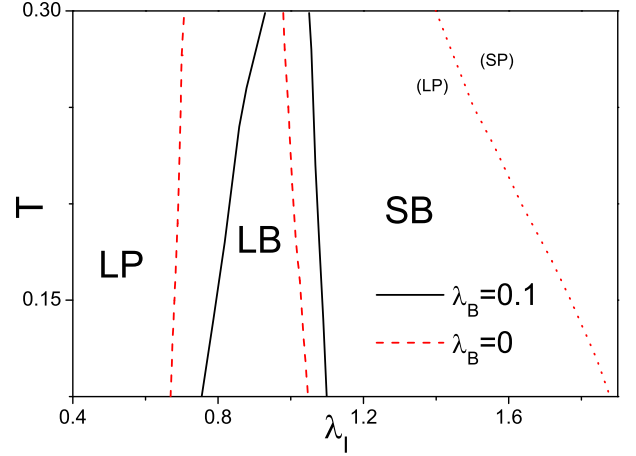}
\caption{Phase diagram of the bipolaron in the plane temperature (in units of $t$) vs. interface el-ph coupling $\lambda_I$  for $\lambda_B=0$ (dash line) and
$\lambda_B=0.1$ (solid line). $LP$ stands for large polaron, $LB$ for large bipolaron, $SB$ for small bipolaron, and $SP$ for small polaron.
The dotted line marks the crossover between large (LP) and small polaron (SP) for $\lambda_B=0$ in the limit of a single particle.}
\label{spectral2}
\end{figure}

In Fig. \ref{spectral2}, we show the phase diagram of the bipolaron for $\lambda_{B}=0$ and $\lambda_{B}=0.1$ in the plane $\lambda_{I}$ vs. temperature. We notice that, in the limit of a single particle, the crossover (dotted line) from LP to small polaron (SP) occurs for high values of  $\lambda_{I}$ in agreement with previous studies. \cite{giulio,single} We stress that the crossover line from LP to LB is shifted towards larger $\lambda_{I}$. The crossover line between LB and SP is less slightly shifted, but always towards larger interface couplings.

\begin{figure}[htb]
\centering
\onefigure[height=0.37\textwidth,angle=0]{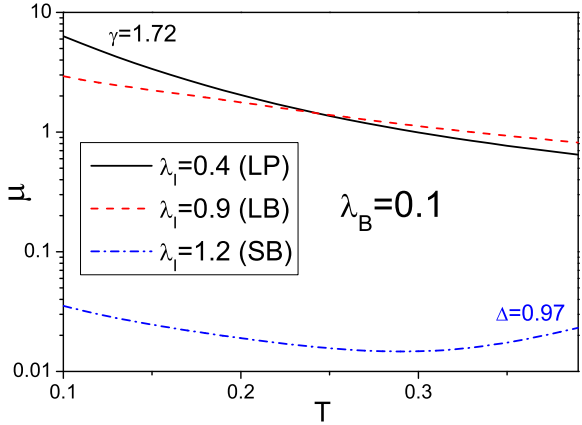}
\caption{The mobility $\mu$ (in units of $\mu_0 = e a^2 / \hbar$) as a function of the temperature (in units of $t$) for different strengths of the interface el-ph coupling: $\lambda_{I}=0.4$ for the large polaron (LP), $\lambda_{I}=0.9$ for the large bipolaron (LB), $\lambda_{I}=1.2$ for the small bipolaron (SP).
The quantity $\gamma$ is the exponent of the power-law $T^{-\gamma}$, while $\Delta$ is the activation energy for the hopping law $\exp{(-\Delta/T)}$.}
\label{density1}
\end{figure}

The main result of this study is related to the analysis of the transport properties. As shown in Fig. \ref{density1}, the mobility has been calculated for different values of $\lambda_I$ at $\lambda_B=0.1$ focusing on the regimes of $LP$, $LB$, and $SB$. For  $\lambda_I=0.4$ (LP), the mobility $\mu$ scales as a power-law ($\mu \sim T^{-\gamma}$, with $\gamma= 1.72$). This behavior is due to the electron diffusion by bulk modes. \cite{vittoriocheck} The main effect of the interface el-ph coupling is to reduce the exponent $\gamma$ starting from values close to $\gamma \simeq 2$. For these regimes of parameters, the role of the interface coupling is similar to that of bulk intra-molecular local modes investigated in a recent paper. \cite{perroni1}

For $\lambda_I=0.9$, we observe a crossover between two regimes. Below $T=0.25$ (see Fig. \ref{spectral1}), the LB is stable. We still find a power law ($\mu \sim T^{-\gamma}$) as in the case of LP, but with the reduced exponent $\gamma \simeq 0.7$. Above $T=0.25$, the mobility still decreases with temperature, but the exponent $\gamma$ is further reduced. Actually, the LP is on the verge of the crossover to SP which occurs at temperatures much larger than
$0.3 t$. Our findings for intermediate coupling regime are relevant for the experimental situation in which the gate is poorly polarizable. \cite{stassen,nature,ono,xia}.

The mobility in SB regime is shown in Fig. \ref{density1} for $\lambda_I=1.2$. It is strongly reduced and changes from a tiny metallic behavior at low temperatures to an insulating activated regime at high temperatures. Actually, at high $T$, $\mu$ goes as $\exp{(-\Delta/T)}$, with
$\Delta \simeq 0.97 t$. Therefore, the bipolaron mobility naturally justifies activation gap of the order the hopping electron energy $t$ for realistic values of the interface el-ph coupling. We stress that, in the limit of a single particle, \cite{single} for $\lambda_I=1.3$, $\Delta$ is of the order of $0.25 t$, and, for the unrealistic value $\lambda_I=2.1$, $\Delta$ is only about $0.41 t$. Finally, we note that the order of magnitude of the mobility is comparable with experimental data in rubrene OFETs. \cite{nature} Actually, the mobility is expressed in natural units, that is in terms of $\mu_0 = e a^2 / \hbar \simeq 7 cm^2 /(V \cdot s) $, taking $a=7 \AA$. \cite{corop}. Therefore, the mobility is about $0.1$ $cm^2 /(V \cdot s)$ at room temperature, that is two orders of magnitude less than that of OFETs exposed to vacuum.

{\it Conclusions and discussions}
In summary, OS gated with polarizable dielectrics provide conditions favorable for the el-ph induced bipolaron formation.
We have examined a model which includes both the bulk el-ph coupling and the long-range interaction of the carriers with the gate modes. The crossover to small bipolarons occurs for realistic el-ph coupling strengths much smaller than those for the small polaron formation. However, within the small bipolaron regime, the mobility is characterized by an activated behavior with gaps much larger than those obtained in the small polaron regime. The results have been discussed in connection with rubrene field-effect transistors.

In this work, we have assumed a simpler interface phonon spectrum and el-ph coupling, in comparison with models in high dimensions. \cite{bussac} However, the interface el-ph coupling shows a long-range behavior similar to that in the more realistic two-dimensonal case. {\it Actually, an important role is played by interface el-ph coupling: in fact, $\lambda_{Inter}$ is of the order of or larger than $1$, while $\lambda_{Bulk}$ is of the order of $0.1$. The features due to bulk coupling are limited to small changes in the bipolaron phase diagrams and the tiny coherent contribution to mobility at very low temperatures.} 

This work is focused on the regime of low density since there are a few examples of highly doped OFET on solid substrates. \cite{RMP,fratini} With increasing density, in the regime of strong el-ph interaction, the SB phase is found to be stable upon clustering. \cite{alexandrov} However, in the intermediate regime, the situation can change upon slightly increasing doping. Actually, a clustering of polarons or bipolarons could take place, and there should also be the possibility of a phase separation between very large undoped regions with large lattice deformations and tiny regions at substantial filling with small deformations. Therefore, the increase of density involves peculiar many-body effects which require deep investigations.

In addition to transport properties, the possibility to distinguish between polaron and bipolaron formation could come from optical spectroscopy and electron spin resonance experiments. \cite{alexandrov} However, there are not so many experiments with single crystal rubrene on highly polarizable dielectric gates. To our knowledge, the only measurements of optical properties of rubrene have been performed with gates made of the poorly polarizable parylene \cite{basov} where polaron/bipolaron signatures do not emerge. Similar situation occurs for experiments based on electron spin resonance. \cite{hasegawa} Therefore, the bipolaron proposal is expected to be more firmly confirmed by different probes not only in rubrene but also in other OS.


\begin{thebibliography}{0}
\bibitem{RMP}
 \Name{T. C. H. Ahn, {\it et al.}}
 \REVIEW{Rev. Mod. Phys.}{78}{2006}{1185}.

\bibitem{hasegawa}
 \Name{T. Hasegawa, {\it et al.}}
 \REVIEW{Sci. Technol. Adv. Mater.}{10}{2009}{24314}.

\bibitem{morpurgo}
 \Name{M. E. Gershenson, {\it et al.}}
 \REVIEW{Rev. Mod. Phys.}{78}{2006}{973}.

\bibitem{stassen}
 \Name{A. F. Stassen, {\it et al.}}
 \REVIEW{App. Phys. Lett.} {85}{2004}{3899}.

\bibitem{nature}
 \Name{I. N. Hulea, {\it et al.}}
 \REVIEW{Nature Mater.} {5}{2006}{982}.

\bibitem{ono}
 \Name{S. Ono, {\it et al.}}
 \REVIEW{App. Phys. Lett.} {94}{2010}{063301}.

\bibitem{cheng}
 \Name{Y. C. Cheng, {\it et al.}}
 \REVIEW{J. Chem. Phys.} {118}{2003}{3764}.

\bibitem{corop}
 \Name{V. Coropceanu, {\it et al.}}
 \REVIEW{Chem. Rev.} {107}{2007}{926}.

\bibitem{hannewald}
 \Name{F. Ortmann, {\it et al.}}
 \REVIEW{New J. Phys.} {12}{2010}{023011}.

\bibitem{troisi_prl}
 \Name{A. Troisi, {\it et al.}}
 \REVIEW{Phys. Rev. Lett.} {96}{086601}{2006}.

\bibitem{vittoriocheck}
 \Name{V. Cataudella, {\it et al.}}
 \REVIEW{Phys. Rev. B} {83}{165203}{2011}.

\bibitem{fratini}
 \Name{S. Fratini, {\it et al.}}
 \REVIEW{New J. Phys.}{10}{2008}{033031}

\bibitem{bussac}
 \Name{N. Kirova, {\it et al.} }
 \REVIEW{Phys. Rev. B}{68}{2003}{235312}.

\bibitem{bussac2}
 \Name{N. Kirova}
 \REVIEW{IJHSES}{17}{2007}{172}.

\bibitem{bussac1}
 \Name{S. J. Konezny, {\it et al.}}
 \REVIEW{Phys. Rev. B}{81}{2010}{045313}.

\bibitem{giulio}
 \Name{G. De Filippis, {\it et al.}}
 \REVIEW{Phys. Rev. B}{82}{2010}{205306}.

\bibitem{single}
 \Name{C. A. Perroni, {\it et al.}}
 \REVIEW{Phys. Rev. B}{85}{2012}{155205}.

\bibitem{liu}
 \Name{C. Liu, {\it et al.}}
 \REVIEW{Adv. Mater.}{23}{2011}{523}.

\bibitem{geskin}
 \Name{V. M. Geskin, {\it et al.}}
 \REVIEW{Chem. Phys. Chem.}{4}{2003}{498}.

\bibitem{bobbert}
 \Name{P. A. Bobbert, {\it et al.}}
 \REVIEW{Phys. Rev. Lett.}{99}{2007}{216801}.

\bibitem{yaping}
 \Name{L. Yaping and J. B. Lagowski}
 \REVIEW{Opt. Mater.}{32}{2010}{1177}.

\bibitem{girlando}
 \Name{A. Girlando, {\it et al.} }
 \REVIEW{Phys. Rev. B} {82}{2010}{035208}.

\bibitem{troisi_adv}
 \Name{A. Troisi}
 \REVIEW{Adv. Mater.} {19}{2007}{2000}.

\bibitem{newus}
 \Name{F. Gargiulo, {\it et al.}}
 \REVIEW{Phys. Rev. B}{84}{2011}{245204}.

\bibitem{SSH}
 \Name{W. P. Su, {\it et al.}}
 \REVIEW{Phys. Rev. Lett.}{42}{1979}{1698}.

\bibitem{nota}
For the calculation of dynamic quantities,
an additional small broadening $\Gamma=0.05 t$ is introduced simulating the effect of a tiny disorder.


\bibitem{lang}
 \Name{I. J. Lang, {\it et al.}}
 \REVIEW{Sov. Phys. JETP}{16}{1963}{1301}.

\bibitem{alexandrov}
 \Name{A. S. Alexandrov and J. T. Devreese}
 \Book{Advances in Polaron Physics}
 \Editor{Springer-Verlag}
 \Publ{Berlin Heidelberg}
 \Year{2010}.

\bibitem{vitmanga}
 \Name{V. Cataudella, {\it et al.}}
 \REVIEW{Phys. Rev. B}{70}{2004}{193105}.

\bibitem{mahan}
 \Name{G.D. Mahan}
 \Book{Many-particle Physics 2nd ed.}
 \Editor{Plenum Press}
 \Publ{New York}
 \Year{1990}.

\bibitem{meSSH}
 \Name{C. A. Perroni, {\it et al.}}
 \REVIEW{Phys. Rev. B}{69}{2004}{174301}.

\bibitem{perroni1}
 \Name{C. A. Perroni, {\it et al.}}
 \REVIEW{Phys. Rev. B}{84}{2011}{014303}.

\bibitem{xia}
 \Name{Y. Xia, {\it et al.}}
 \REVIEW{Phys. Rev. Lett.}{105}{2010}{036802}.

\bibitem{basov}
 \Name{Z. Q. Li, {\it et al.}}
 \REVIEW{Phys. Rev. Lett.}{99}{2007}{016403}.



\end{thebibliography}
\end{document}